\documentstyle[epsf]{mn}
 \input epsf.sty
\newif\ifAMStwofonts
\ifoldfss

  \ifCUPmtlplainloaded \else
    \NewTextAlphabet{textbfit} {cmbxti10} {}
    \NewTextAlphabet{textbfss} {cmssbx10} {}
    \NewMathAlphabet{mathbfit} {cmbxti10} {} 
    \NewMathAlphabet{mathbfss} {cmssbx10} {} 
  \fi
  \ifAMStwofonts
    \ifCUPmtlplainloaded \else
      \NewSymbolFont{upmath} {eurm10}
      \NewSymbolFont{AMSa} {msam10}
      \NewMathSymbol{\upi}     {0}{upmath}{19}
      \NewMathSymbol{\umu}     {0}{upmath}{16}
      \NewMathSymbol{\upartial}{0}{upmath}{40}
      \NewMathSymbol{\leqslant}{3}{AMSa}{36}
      \NewMathSymbol{\geqslant}{3}{AMSa}{3E}

       \let\le=\leqslant
       \let\ge=\geqslant
    \fi
  \fi
\fi 

\ifnfssone
  \newmathalphabet{\mathit}
  \addtoversion{normal}{\mathit}{cmr}{m}{it}
  \addtoversion{bold}{\mathit}{cmr}{bx}{it}
  \newmathalphabet{\mathbfit} 
  \addtoversion{normal}{\mathbfit}{cmr}{bx}{it}
  \addtoversion{bold}{\mathbfit}{cmr}{bx}{it}
  \newmathalphabet{\mathbfss} 
  \addtoversion{normal}{\mathbfss}{cmss}{bx}{n}
  \addtoversion{bold}{\mathbfss}{cmss}{bx}{n}
  \ifAMStwofonts
    \ifCUPmtlplainloaded \else
      \UseAMStwoboldmath
      \makeatletter
      \new@mathgroup\upmath@group
      \define@mathgroup\mv@normal\upmath@group{eur}{m}{n}
      \define@mathgroup\mv@bold\upmath@group{eur}{b}{n}
      \edef\UPM{\hexnumber\upmath@group}
      \new@mathgroup\amsa@group
      \define@mathgroup\mv@normal\amsa@group{msa}{m}{n}
      \define@mathgroup\mv@bold\amsa@group{msa}{m}{n}
      \edef\AMSa{\hexnumber\amsa@group}
      \makeatother
      \mathchardef\upi="0\UPM19
      \mathchardef\umu="0\UPM16
      \mathchardef\upartial="0\UPM40
      \mathchardef\leqslant="3\AMSa36
      \mathchardef\geqslant="3\AMSa3E

       \let\le=\leqslant
       \let\ge=\geqslant
    \fi
  \fi
\fi 

\ifnfsstwo
  \DeclareMathAlphabet{\mathbfit}{OT1}{cmr}{bx}{it}
  \SetMathAlphabet\mathbfit{bold}{OT1}{cmr}{bx}{it}
  \DeclareMathAlphabet{\mathbfss}{OT1}{cmss}{bx}{n}
  \SetMathAlphabet\mathbfss{bold}{OT1}{cmss}{bx}{n}
  \ifAMStwofonts
    \ifCUPmtlplainloaded \else
      \DeclareSymbolFont{UPM}{U}{eur}{m}{n}
      \SetSymbolFont{UPM}{bold}{U}{eur}{b}{n}
      \DeclareSymbolFont{AMSa}{U}{msa}{m}{n}
      \DeclareMathSymbol{\upi}{0}{UPM}{"19}
      \DeclareMathSymbol{\umu}{0}{UPM}{"16}
      \DeclareMathSymbol{\upartial}{0}{UPM}{"40}
      \DeclareMathSymbol{\leqslant}{3}{AMSa}{"36}
      \DeclareMathSymbol{\geqslant}{3}{AMSa}{"3E}

       \let\le=\leqslant
       \let\ge=\geqslant
    \fi
  \fi
\fi 
\ifCUPmtlplainloaded \else
  \ifAMStwofonts \else 
    \def\upi{\pi}
    \def\umu{\mu}
    \def\upartial{\partial}
  \fi
\fi

\title{Nova Sagittarii 1998 (V4633~Sgr) -- A Permanent Superhump
  System or an Asynchronous Polar?}

\author[Y. Lipkin, E.M. Leibowitz, A. Retter, O. Shemmer]
   {Y.~Lipkin$^1$,
    E.M.~Leibowitz$^1$,
    A.~Retter$^{2,3}$,
    O. Shemmer$^1$\\
 $^1$School of Physics and Astronomy and the Wise Observatory,
Raymond and Beverly Sackler Faculty of Exact Sciences,\\
Tel-Aviv University, Tel Aviv, 69978, Israel; yiftah@wise.tau.ac.il,
elia@wise.tau.ac.il, ohad@wise.tau.ac.il\\
 $^2$Dept. of Physics, Keele University, Keele, Staffordshire,
ST5 5BG, U.K.; ar@astro.keele.ac.uk;\\
$^3$School of Physics, University of Sydney, 2006, Australia;
retter@physics.usyd.edu.au\\}
\pagerange{\pageref{firstpage}--\pageref{lastpage}}
\pubyear{2001}

\def\LaTeX{L\kern-.36em\raise.3ex\hbox{a}\kern-.15em
    T\kern-.1667em\lower.7ex\hbox{E}\kern-.125emX}

\begin{document}
\def\gtorder{\mathrel{\raise.3ex\hbox{$>$}\mkern-14mu
             \lower0.6ex\hbox{$\sim$}}}
\def\ltorder{\mathrel{\raise.3ex\hbox{$<$}\mkern-14mu
             \lower0.6ex\hbox{$\sim$}}}

\label{firstpage}

\maketitle

\begin{abstract}

We report the results of observations of V4633~Sgr (Nova
Sagittarii 1998) during 1998-2000.
Two photometric periodicities were present in the light curve during
the three years of observations: a stable one at $P=3.014$~h,
which is probably the orbital period of the underlying binary system,
and a second one of lower coherence, approximately $2.5$ per cent longer than
the former.
The latter periodicity may be a permanent superhump, or alternatively,
the spin period of the white dwarf in a nearly synchronous magnetic
system.
A third period, at $P=5.06$~d, corresponding to the beat between the
two periods was probably present in 1999.
Our results suggest that a process of mass transfer took place in the
binary system since no later than two and a half months after the nova
eruption.
We derive an interstellar reddening of $E(B-V)\sim0.21$ from our
spectroscopic measurements and published photometric data, and
estimate a distance of $d\sim 9$ kpc to this nova.
\end{abstract}


\section{Introduction}
Nova Sgr~1998 was discovered on 1998 March 22 by Liller
(1998). Brightest visual magnitude of $7.4$~mag was reported by Jones
(1998) on March $23.7$.
Liller \& Jones (1999) classified V4633~Sgr as a fast nova, with
$t_{3}\approx 35$~d for the visual observations, and $\approx48$~d in 
CCD broadband $V$.
An early spectrum of V4633~Sgr revealed slow expansion
velocities and massive presence of iron, implying a Fe\,{\sc ii}
classification (Della Valle, Pizzella \& Bernardi 1998).

Skiff (1998) reported no definite object at the location of V4633~Sgr in
the Palomar Sky Survey, setting a lower limit of $12$~mag on the
outburst amplitude.

Spectropolarimetry of V4633~Sgr shortly after maximum brightness did
not yield evidence for intrinsic polarization (Ikeda, Kawabata \&
Akitay 2000).

IR-spectrophotometry indicated that V4633~Sgr was in early stages of
its coronal phase in 1999 August (Rudy et al. 1999),
and revealed strong coronal lines, and a relatively low reddening in
2000 July (Rudy et al. 2000).

Lipkin, Retter \& Leibowitz (1998) reported a photometric modulation
in the light curve (LC) of V4633~Sgr, with a period of $0.17330$ or
$0.14765\pm 0.00011$~d, which are 1-day aliases of each other.
The modulation was detected eleven weeks, and possibly as early as six
weeks after the eruption.
Later on, Lipkin \& Leibowitz (2000) found that another 1-day alias,
at $0.128791$~d, is in fact
the dominant periodicity in the LC. They also reported the
discovery of a second photometric periodicity at $0.125573$~d,
modulating the brightness of the star along with the first one during
1999 and also in 1998.

In this paper we describe in detail the photometric properties of the
V4633~Sgr during the 1998-2000 seasons.
We also report on a few spectroscopic observations that we performed
on this star, and on implications of these data on some properties of
this system.

\section{Observations and data reduction}
\subsection{Photometry}
\label{SecPhot}
We performed photometry of V4633~Sgr during 34 nights in 1998, 36
nights in 1999, and 26 nights in 2000,
using the Tektronix 1K back-illuminated CCD, mounted on the 1-m
telescope at the Wise Observatory (WO).
Details on the telescope and instrument are given by Kaspi et al. (1995).

Photometry was conducted either through $I$ filter,
or switching sequentially between $I$ and $V$, or between $I$, $V$, and $B$
filters.
Logs of the observations are given in appendix A.

Photometric measurements on the bias-subtracted and flat-field
corrected images were
performed using the NOAO IRAF\footnote{IRAF (Image Reduction and Analysis
Facility) is distributed by the National Optical Astronomy Observatories,
which are operated by AURA, Inc., under cooperative agreement with the National
Science Foundation.} {\sc{daophot}} package (Stetson 1987).
Instrumental magnitudes of V4633~Sgr, as well as of a few dozens of
reference stars, depending on image quality, were obtained for each frame.
A set of internally consistent nova magnitudes was obtained using the
WO reduction program {\sc{daostat}} (Netzer et al. 1996).
Good seeing conditions on 1998 September 19 were used to calibrate
the magnitudes of V4633~Sgr, as well as of about a dozen nearby
comparison stars.
We used the calibrated comparison stars to convert all the
measurements of V4633~Sgr into calibrated magnitudes.

In our program we obtained 84 nights of continuous time-series, accumulating
a total of 8250 data-points in $I$, 2392 in $V$ and 756 in $B$.

On 2000 August 4, 20, and 21, we observed V4633~Sgr in the 'fast
photometry' mode (Leibowitz, Ibbetson \& Ofek 1999). 
On the first night we observed for 2.5~h, with time-resolution
of 10~s, using no filter ('clear').
On the other two nights, we observed through $I$ filter, with time-
resolution of 20~s.
The data were reduced in the manner described above.

\subsection{Spectroscopy}
V4633~Sgr was observed spectroscopically at WO on four nights:
1998 July 5 and August 30, and 1999 May 2 and July 6. The spectra
were taken with the WO Faint Object Spectrograph
and Camera (FOSC) described in Brosch \& Goldberg (1994), and operated
at the f/7 Ritchey -- Chr\`{e}tien focus of the WO 1-m telescope. The
Tektronix 1K CCD was used as the detector.
We applied the method of long-slit spectroscopy
whereby both V4633~Sgr and a bright comparison star were included in the
slit
(see for example Kaspi et al. 2000).
The comparison star used was non-variable to within $\sim$2 per cent.
We used a 10''-wide slit along with a 600 line mm$^{-1}$ grism,
yielding a dispersion of 4~\AA/pix ($\sim$8~\AA\ resolution).
On the first two nights the spectrograph was set to
cover the spectral range $\sim$3600 -- 7200~\AA, while in the last two nights we
covered the range $\sim$4000 -- 7800~\AA.
Two exposures of the spectrum of the nova were taken on each night.

Reduction of the bias-subtracted and flat-field corrected spectra was
carried out in the usual manner using IRAF with its {\sc{specred}} and
{\sc{onedspec}} packages. 
The spectra were dispersion corrected using
a He-Ar arc spectrum, which was taken on each night in between the pair
of the nova spectra.
Each spectrum of the nova was divided by the spectrum of the comparison star
observed simultaneously through the same slit. The two sets of nova/star
spectrum ratios obtained on each night were compared to each other and were
found to differ by no more than $\sim$10 per cent. The average of the two
ratios was then taken as the representative ratio 
for that night.
The spectra were calibrated to an absolute
flux scale by multiplying each mean nova/star ratio by a flux calibrated
spectrum of the comparison star. This spectrum, in turn, was flux calibrated
using the WO standard sensitivity function and extinction curve. These
do not change appreciably from night to night, and they are updated from time to
time at WO using spectrophotometric standard stars. The absolute flux
calibration has an uncertainty of $\sim$10 per cent, but the relative flux
uncertainties within each spectrum are of order $2-3$ per cent.

\section{Data analysis}
\label{SecLC}
\begin{figure}
\centerline{\epsfxsize=3.5in\epsfbox{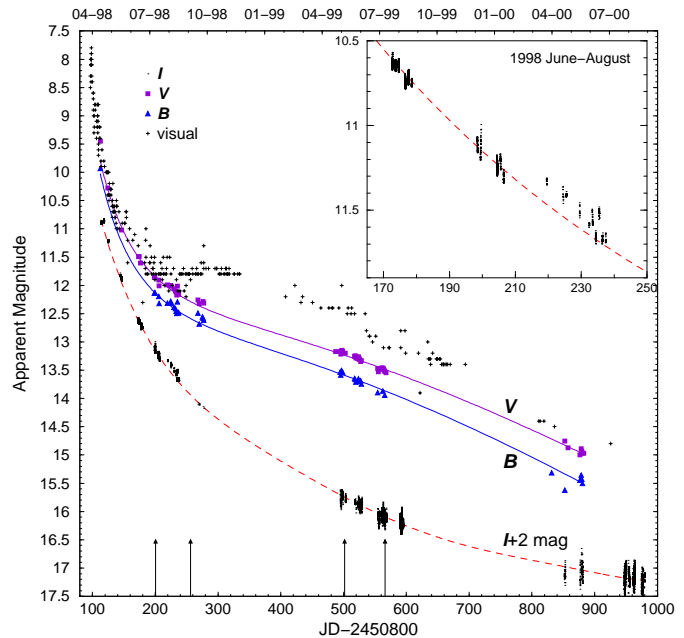}}
 \caption{Light curves of V4633~Sgr. The points in the $B$ and $V$ LCs
   are mean magnitudes of each night of observations. The $I$ LC
   comprises of all the data points compiled in our observations. 
   For convenience, the $I$ LC was shifted down by $2$~mag. 
   The overlaid curves represent the long-term decline of the nova in
   each photometric band.
   The arrows at the bottom of the figure indicate dates of
   spectroscopic observations.
   The scale on the top axis is the date, in MM-YY format.
   The inset figure is a zoom-in view of the $I$ LC during 1998
   June -- August, showing the 'bump' in the LC in August.}
\label{figLc}
\end{figure}

Light curves of V4633~Sgr from discovery to 2000 July are presented
in Fig.~\ref{figLc}. The visual LC was compiled using data taken
from VSNET\footnote{VSNET--Variable Stars Network, Kyoto, Japan\\
URL: http://www.kusastro.kyoto-u.ac.jp/vsnet/}.
The $I$, $V$, and $B$ LCs were compiled using data obtained in our
program.
Note that the apparent small vertical lines in the $I$ LC are not
error bars but dense individual successive points observed in a single
night. Bars representing the observational errors in our measurements
are below the resolution limit of this figure.

The visual and $V$ LCs show an apparent change in the slope, becoming
more moderate about three months after maximum light
(Fig.~\ref{figLc}). 
Most of the 1998 photometry was conducted around the time the slope
changed.
Shortly after, in 1998 July -- August, the brightness of the star
deviated systematically from the long-term trend given by the fitted
curve, forming an apparent bump in the LC (Fig.~\ref{figLc}, inset frame).

\begin{figure*}
  \centerline{\epsfxsize=5.5in\epsfbox{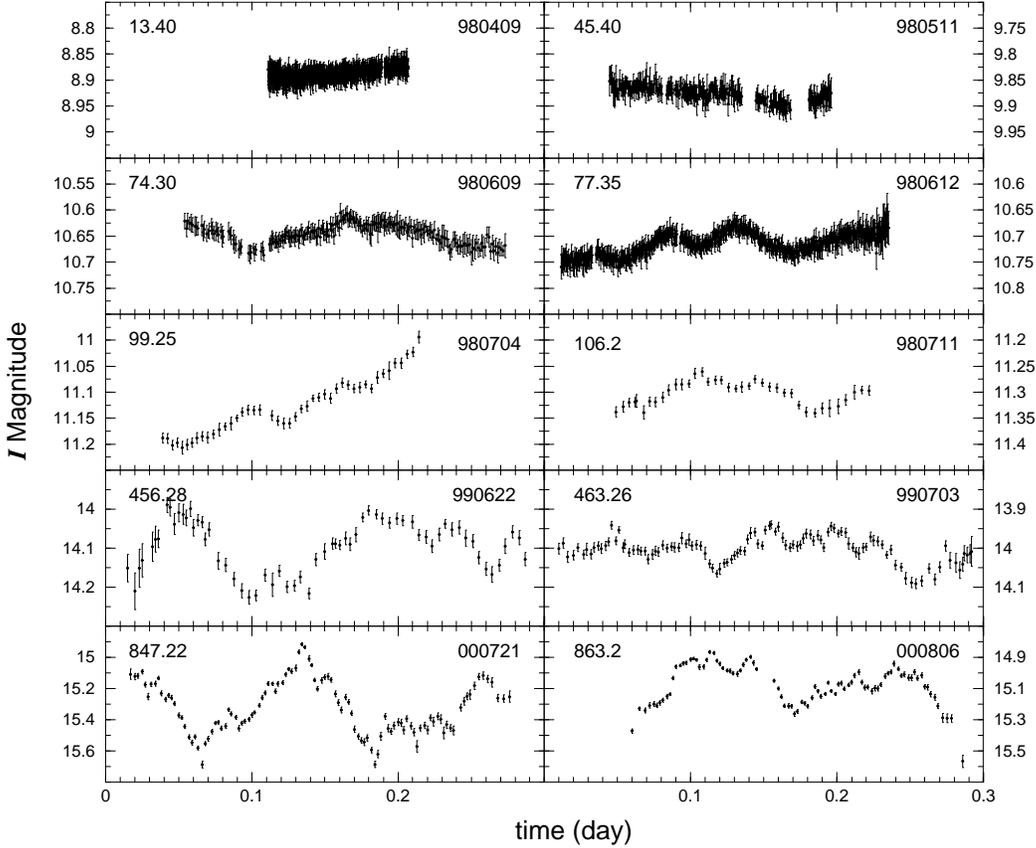}}
  \caption{A Sample of nightly light curves of V4633~Sgr from different
    epochs. On the left of each frame is the time (HJD-2450900) of the
    beginning of observation. The date of observation is on the right
    of each frame, in YYMMDD format. Note the difference in the Y-axis
    scales.}
\label{figLCs}
 \end{figure*}

A panel of sample $I$-band LCs from different epochs is shown in
Fig.~\ref{figLCs}.
Nightly LCs show almost no visible variation until 1998 May.
Fragmented LCs in May show some variation, while in June modulations
on a time-scale of $\sim3$~h are clearly visible.
In July and August, the variations took other forms. In a few nights
the variations are quasi-periodic
but on a somewhat different time-scale than in June. 
On few other nights, the brightness of the star varied monotonically
during the entire nightly run.
In all our subsequent observations in 1999 and 2000, the variations
returned to the oscillation mode of 1998 June, albeit with an ever increasing
amplitude.

\subsection{The 1999 light curve}
\label{SecLC99}
\begin{figure*}
  \centerline{\epsfxsize=5.5in\epsfbox{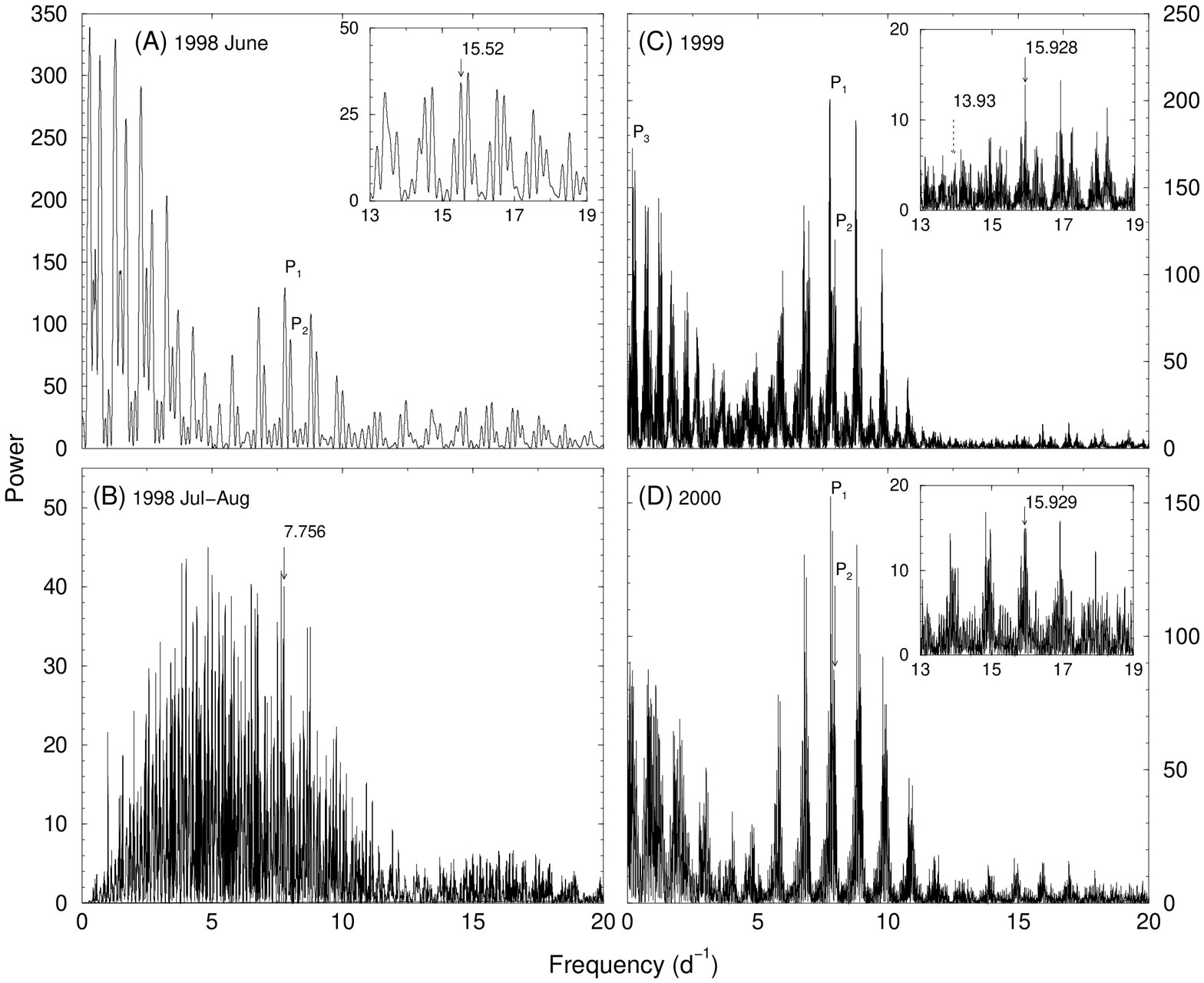}}
  \caption{Normalized power spectrum of four datasets. The peaks
    corresponding to $P_1$ and $P_2$ are marked in each PS.\newline
    {\bf A.} The 1998 June $I$ dataset. The peak marked in the inset
    frame probably corresponds to the first overtone of $P_1$.\newline
    {\bf B.} The 1998 July -- August $V$ dataset. A night of monotonic
    trend was excluded. The data was pre-whitened by subtracting the
    mean magnitude from each night. An arrow marks a peak at
    $7.756$~d$^{-1}$, which is the seventh highest peak in the PS.\newline
    {\bf C.} The 1999 $I$ dataset. The low end of the PS is dominated
    by $P_3$, at 0.1976~d$^{-1}$. The first overtone of $P_2$, at
    15.928~d$^{-1}$ is marked in the inset figure.
    The first overtone of the 6.963~d$^{-1}$ 1-day alias
    of $P_2$ is expected at 13.926~d$^{-1}$ (marked by a dashed arrow
    in the inset figure). However, no noticable signal is detected in
    the vicinity of this frequency.\newline
    {\bf D.} The 2000 $I$ dataset. The first overtone of $P_2$ is marked
    in the inset figure.}
\label{figPs}
 \end{figure*}

We first discuss the data of 1999 since this season is better sampled
than the other two.
Fig.~\ref{figPs}-C shows the normalized power spectrum (PS) (Scargle
1982) of our 1999 $I$-band data, after eliminating the long term
decline by subtracting a fourth-degree polynomial from the 1998-1999
LC.

The PS is dominated by two similar alias patterns around two central
frequencies, $7.7660$~d$^{-1}$, and $7.9628$~d$^{-1}$,
corresponding to the periodicities $0.128760\pm0.000013$~d (hereafter
$P_1$), and $0.125589\pm0.000016$~d ($P_2$).

To derive the quoted periods, we performed a grid search in the
$\chi^2$ space, fitting to the data a polynomial term representing
the secular decline of the nova and a pair of periods near the values
of $P_1$ and $P_2$ obtained from the PS. The grid was then examined to
find the pair of periods yielding the lowest value in the $\chi^2$ space.

The errors of the two periods correspond to a 1-$\sigma$ confidence
level, and were derived by a sample of 2000 Bootstrap simulations
(Efron \& Tibshirani 1993).

We used the tests described in Retter, Leibowitz \& Kovo-Kariti (1998)
to confirm the independence of the two periodicities.
Similar results were also obtained from the PSs of our 1999 $V$ and
$B$ datasets.

At the right-hand side of Fig.~\ref{figPs}-C, the first overtone of $P_2$ is
detected at $15.928$~d$^{-1}$, well above the noise level in its
vicinity. Such a feature is expected, due to the asymmetric shape of
the signal (see Sec.~\ref{SecFold}).

The lower end of the 1999 PS (Fig.~\ref{figPs}-C) is dominated by a
structure of interdependent peaks, the highest of which, designated $P_3$, is at
$0.1976$~d$^{-1}$ (5.06~d) with a full amplitude of $0.096$~mag.
This periodicity corresponds to the beat period between $P_1$
and $P_2$. 
The signal was found to be independent from $P_1$ and $P_2$. 
It was not detected in our $V$ and $B$ LCs. However, these
datasets are of lower quality than the $I$ dataset, and span a shorter
time.
Due to relatively high noise of the PS near $P_3$ and the
fragmented nature of the LC on time-scales of a few days, the
reliability of $P_3$ should be addressed with some caution, until
it is confirmed by further observations.

In the 1999 $I$-band PS, the 1-day alias of $P_2$, at $6.961$~d$^{-1}$ is
stronger than the peak associated with $P_2$ (Fig.~\ref{figPs}-C).
The same result occurred in few other tests we conducted, for various
subsets of the data, as well as in the different bands and datasets,
and using various de-trending methods. Similarly, in a small number of
tests the signal at $6.76$~d$^{-1}$, or the one at $8.76$~d$^{-1}$,
dominated the alias structure of $P_1$, rather than the one at
$7.76$~d$^{-1}$.

These results introduce some uncertainty to our selection of
$7.76$~d$^{-1}$ and $7.96$~d$^{-1}$ for $P_1$ an $P_2$, respectively.
However, we consider this selection firm, due to the dominance of
these periods in the bulk of our tests.
Further support to this selection comes from the presence in the PS of the
first overtone of $7.96$~d$^{-1}$, and the absence of any noticeable
signal at the frequency of the expected first
overtone of $6.96$~d$^{-1}$ (Fig.~\ref{figPs}-C, inset figure).
The presence in the PS of $P_3$ -- the beat of $7.76$~d$^{-1}$ and
$7.96$~d$^{-1}$, is yet another strong argument for selecting these two
periods.

\subsection{The 1998 May -- June light curve}
The PS of the $I$-band data obtained during six nights in 1998 June
(Fig.~\ref{figPs}-A) resembles that of 1999. Two peaks
at $7.782$ and $7.999$~d$^{-1}$, each of which is the centre of a 1, 1/2,
1/3 (etc.)-day alias pattern, dominate the PS.
The group of peaks at the lower end of the PS are of questionable
reliability due to the short time span of the dataset, and since
they are sensitive to the method used to de-trend the strongly
declining LC.

The values of $P_1$ and $P_2$, derived by simultaneously fitting two
periods and a linear term to the LC, are
$0.12893\pm0.00015$~d and $0.12523\pm0.00033$~d.
A peak at $15.52$~d$^{-1}$ (Fig.~\ref{figPs}-A) probably corresponds to
the first overtone of $P_1$, which is expected at this frequency.

Adding the fragmentary time-series obtained in 1998 May to the June
data, the power of the two peaks corresponding to $P_1$ and $P_2$
increased in the combined PS (not shown), relative to the PS of June only.
Indeed, examination of the short LC of May revealed a hump that is in
fair agreement with the modulations of the June LC, extrapolated to the
times of observation in May.
Thus, it is likely that the light of the star was modulated by at
least one of the two periodicities as early as 1998 May.

\subsection{The 1998 July -- August light curve}
\label{SecAug98}
The PS of the $V$-band data gathered during 13 nights in 1998 July -- August
(Fig.~\ref{figPs}-B) is different in its structure and details from
the former two PSs. A broad excess of power in the vicinity of $5$~d$^{-1}$
dominates the PS, but no obviously significant peak stands out above
the wide hump.
Looking for the known periodicities, a peak at $7.756$~d$^{-1}$ is
found (marked with an arrow in Fig.~\ref{figPs}-B).
However, this peak is well within the noise level and there is a high
a-priori probability for its presence in the PS as a result of
random coincidence. The July -- August data are therefore consistent with
a LC which is not significantly modulated by either of the periods
$P_1$ or $P_2$.

To further test the difference between the July -- August data and that of
June, we constructed an artificial LC by extrapolating the signals of
the June LC onto the actual times of observation of the July -- August LC.
Comparing the observed LC and the artificial one, there was only
little resemblance between the two in the phases and shapes
of the modulations.
Also, in contrast to the PS of the actual data (Fig.~\ref{figPs}-B),
the periods of 1998 June were clearly detectable in the
PS of the artificial LC.

\subsection{The 2000 light curve}
\label{SecLC2000}
The PS of the $I$-band data of 2000 (Fig.~\ref{figPs}-D), is dominated
by the signal of $P_1$ at $7.795$~d$^{-1}$.
The signal of $P_2$, at $7.964$~d$^{-1}$, is obscured by the
alias structure of $P_1$, but becomes the dominant feature in the
residual PS once $P_1$ is removed from the data.
A weak signal at $15.929$~d$^{-1}$ is probably the
first overtone of $P_2$.
A simultaneous fit of two periods and a linear term to the
data yields the best-fitting values $P_1=0.128292\pm0.000007$~d, and
$P_2=0.125570\pm0.000010$~d.

Finally, we looked for periodic variations in the data accumulated in the three
2000 nights of fast photometry (Sec.~\ref{SecPhot}). We found no sign
in the data for any periodicity in the range of a few tens of seconds
to a few tens of minutes.

\subsection{Stability of the signals}
\label{SecPdot}
The value of $P_2$ measured in 2000 is just $0.015$ per cent smaller
than in 1999.
The difference amounts to only 1.2\,$\sigma$ of the uncertainty in the
derived value of the periods themselves.
The two values are therefore consistent with the notion that $P_{2}$
is the same in both years.
This is not the case for $P_1$. 
The value measured in 2000 is 0.3 per cent smaller than in 1999,
and the difference is highly significant: 
more than 30\,$\sigma$.

\begin{figure}
  \centerline{\epsfxsize=3.5in\epsfbox{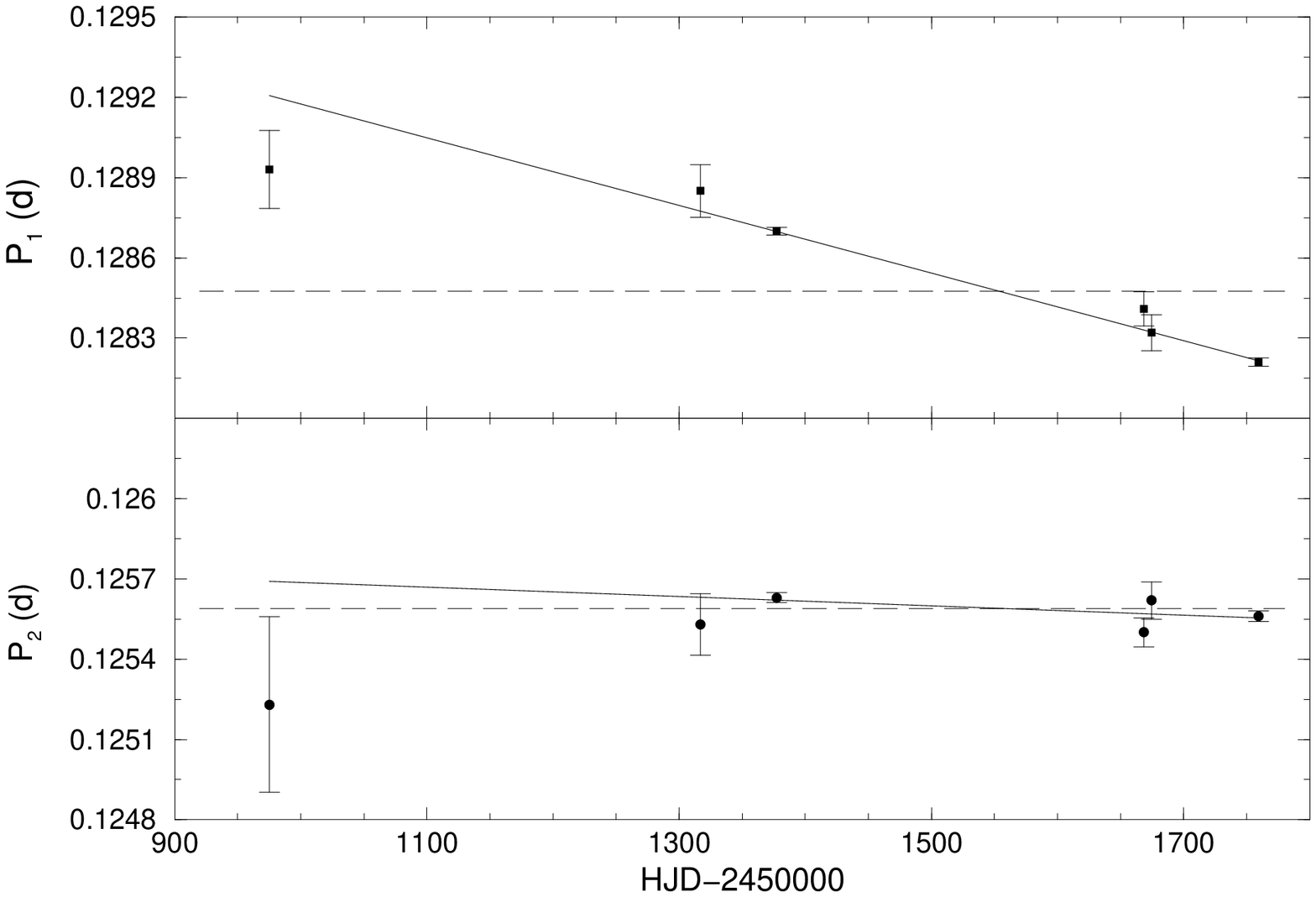}}
  \caption{Variation of $P_1$ (top panel) and $P_2$ (bottom panel) in
    1998-2000. Solid lines are linear fits to the data. Dashed lines
    are the weighted average of the measured periods.}
\label{figPdot}
 \end{figure}

To further examine the stability of the periodicities we measured
$P_1$ and $P_2$ in six different datasets during the 1998-2000 time
interval, in the manner described in Sec.~\ref{SecLC99}.
The measured values of $P_2$ are scattered around the average value of
$0.12559$~d, although a linear fit yields a formal rate of period change 
$\dot P_2=(-1.7\pm0.7)\times 10^{-7}$ (Fig.~\ref{figPdot}, bottom
panel).
We consider this result as consistent with a constant period. The slope
for $P_1$ is highly significant:
$\dot P_1=(-1.26\pm0.05)\times 10^{-6}$ (Fig.~\ref{figPdot}, top
panel). 

\subsection{Waveforms and amplitudes}
\label{SecFold}
\begin{figure}
  \centerline{\epsfxsize=3.5in\epsfbox{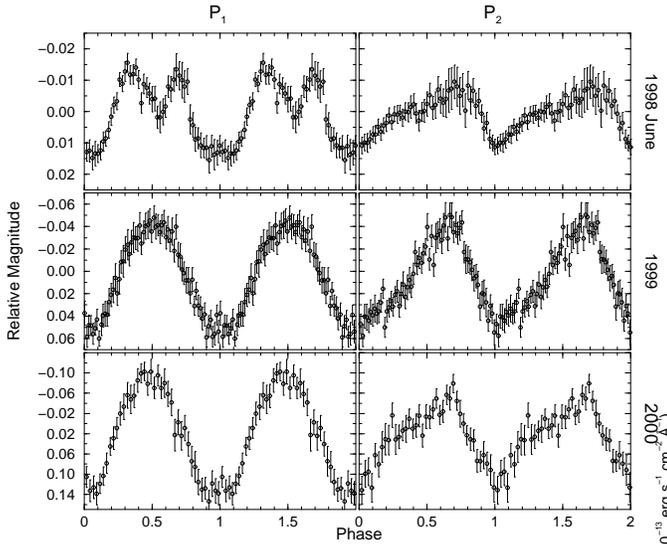}}
  \caption{The $I$ LCs of the three observational seasons folded on the
    period of $P_1$ (left), and on $P_2$ (right). Discrete points are
    mean magnitudes in each of 50 equal bins covering the $0-1$ phase interval.
    The inserted bars are the standard deviation in the value of the mean
    in each bin. Note the difference in the Y-axis scales.}
\label{figFold}
\end{figure}

The waveforms of $P_1$ and $P_2$ in 1998 June, 1999
and 2000 are shown in Fig.~\ref{figFold}. 
In each case, we 'pre-whitened' the LC before folding by removing
the signal of the other periodicity, as well as a polynomial
term representing the decline in the brightness of the nova.
In the 1999 dataset $P_3$ was subtracted as well.

The waveform of $P_1$ was symmetric during the three observational
seasons.
In 1998 June, a clear dip of $0.012$~mag was imposed on the
primary maximum.
In 1999 the waveform transformed into a nearly sinusoidal
shape, which was maintained also in 2000 (Fig.~\ref{figFold}, left
panels).
The peak-to-peak amplitude of $P_1$ was $0.030$~mag in 1998 June, 
$0.105$~mag in 1999, and $0.25$~mag in 2000.

$P_2$ maintained an asymmetric shape during the three observational
seasons, with a slow rise and a fast decline 
(Fig.~\ref{figFold}, right panels).
The peak-to-peak amplitude of $P_2$ was $0.019$~mag in 1998 June,
$0.100$~mag in 1999, and $0.19$~mag in 2000.

The waveforms of $P_1$ and $P_2$ in 1999 in $V$ and $B$, as well as
those obtained from the much limited $V$-band data in 1998 June and in 2000,
were similar to the ones in $I$.

The limited data in $V$ and $B$ do not allow an accurate
tracking of the amplitudes of the two signals.
However, some information on the change in amplitude may be gained by
inspecting the secular change in nightly variation, e.g. by following
the secular change in the standard deviation (STD) of nightly LCs. 
The variation in $I$ steadily increased by about $0.05$~mag~yr$^{-1}$
during 1998-2000, consistent with the increasing amplitudes of the two
periodicities described above. 
The STD of the $V$ magnitudes has not changed significantly in
1998-1999, maintaining a value of $\sim0.027$~mag, and increased in 2000
to $\sim0.073$~mag.
In $B$ it decreased, from $\sim0.040$~mag in 1988, to
$\sim0.023$~mag in 1999.
One should bare in mind that these trends reflect not only changes due
to the brightness variations of sources within the binary system of
V4633~Sgr but also some varying contribution of the nebula to the
total light of the source. Thus, during 1998-1999 the contribution of
the nebula in the $V$-band increased from $40$ per cent to about $70$
per cent (section~\ref{SecSpec}), implying that the amplitude of the
variations in the stellar $V$ continuum was in fact larger than
indicated by the STD values.

\subsection{Spectroscopy}
\label{SecSpec}
\begin{figure}
 \centerline{\epsfxsize=3.351in\epsfbox{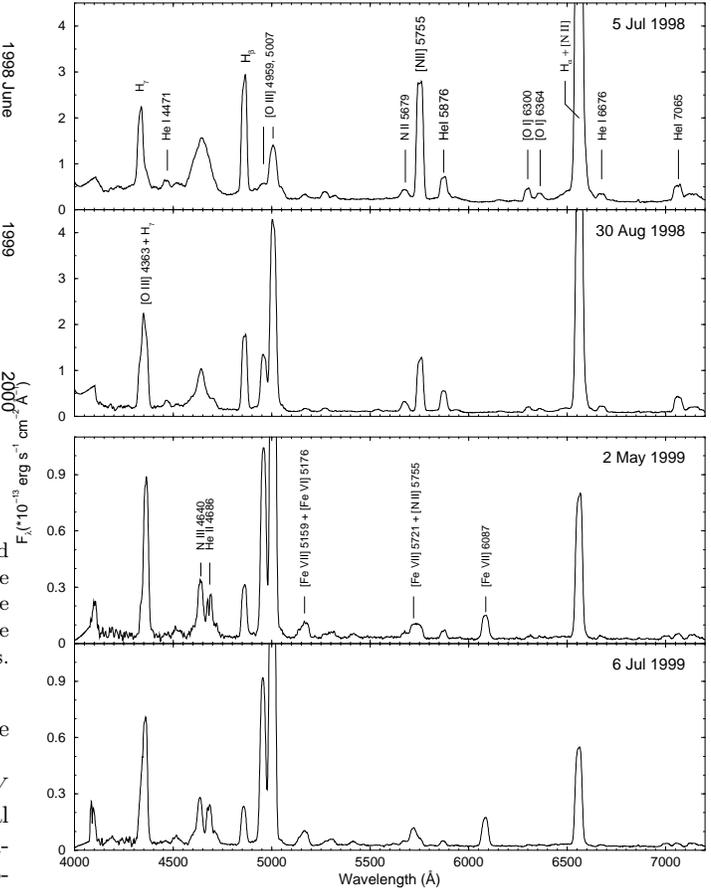}}
  \caption{Spectra of V4633~Sgr obtained at WO. Each emission-line
    identification is given only on the initial appearance of that
    line in the spectrum, and it remains valid for later spectra.}
 \label{figSpec}
 \end{figure}

Four spectra obtained at WO in 1998-1999 are plotted in
Fig.~\ref{figSpec}.
Fluxes of a few of the emission lines are shown in Table~\ref{tableSpec}.
The spectrum of 1998 July 5 features
prominent Balmer lines as well as the strong auroral line
[N\,{\sc ii}]~$\lambda 5755$. The nebular lines
[O\,{\sc iii}]~$\lambda\lambda4959,5007$ are also seen. By 1998 August 30,
[O\,{\sc iii}]~$\lambda\lambda4959,5007$ and [O\,{\sc iii}]~$\lambda4363$ became
stronger, and the Balmer lines and [N\,{\sc ii}]~$\lambda 5755$ faded. 
The two spectra of 1999 were dominated by the
[O\,{\sc{iii}}]~$\lambda\lambda4959,5007$ lines.
The decline of the Balmer lines and [N\,{\sc ii}]~$\lambda 5755$ continued,
while the auroral line [O\,{\sc iii}]~$\lambda 4363$ became more
dominant. Higher ionization lines, of [Fe\,{\sc vii}] appeared, and became
stronger. 
All four spectra are in the auroral phase, according to the Tololo
classification system (Williams et al. 1991, Williams, Phillips
\& Hamuy 1994). The 1998 July 5 spectrum is probably classified $A_n$,
and the other three spectra are probably in the $A_o$ phase.

\begin{table}
  \caption{V4633~Sgr. Intensity of selected emission lines relative to
    H$\beta$. The estimated uncertainties range between 3~per
    cent in the strongest lines, to about 10 per cent in weak lines.}
  \begin{tabular}{@{}||l|c|c|c|c||@{}}
Line                     &\multicolumn{4}{c}{Flux (H$\beta=100$)}\\
Identification                 &  980705    &       980830& 990502& 990706\\
H$\gamma$/[O\,{\sc iii}]~4363 &    86 &     163 &    344  &   424  \\
He\,{\sc i}~4471              &    7.3  &     8.4  &           &          \\
H$\beta$                      &    100   &     100   &    100    &   100    \\
{[O\,{\sc iii}]}~4959         &    4.9  &     73 &    407  &   491  \\
{[O\,{\sc iii}]}~5007         &    58 &     300 &    1246 &   1492 \\
N\,{\sc ii}~5679              &    13.0 &     15.4 &           &          \\
{[N\,{\sc ii}]}~5755          &    141 &     83 &           &          \\
He\,{\sc i}~5876              &    23 &     31 &    18.2  &   14.9  \\
{[Fe\,{\sc vii}]}~6085        &          &           &    64  &   98  \\
{[O\,{\sc i}]}~6300           &    13.1 &     9.1  &           &          \\
{[O\,{\sc i}]}~6364           &    8.8  &     6.5  &           &          \\
H$\alpha$+{[N\,{\sc ii}]}   &    678 &     520 &    381  &   352  \\
He\,{\sc i}~6676              &    5.5  &     9.1 &    7.2   &   5.6   \\
He\,{\sc i}~6065              &    16.2 &     27 &    12.6  &   13.9  \\

\hline
$H\beta$            &    77 &     50 &     7.9  &   5.7   \\
\multicolumn{5}{l}{($\times10^{-13}$erg~s$^{-1}$~cm$^{-2}$)}\\
  \end{tabular}
\label{tableSpec}
\end{table} 

In each of the spectra we calculated the integrated $V$ magnitude of
the star by convolving the observed spectral energy distribution with
the transmission curve of the $V$ filter.
The results agreed with the values obtained from photometry.
The spectra also allowed us to subtract from the integrated $V$
brightness the contribution of the emission lines that originate
mostly in the nebula. As expected, when considered alone, the $V$
continuum faded faster than the integrated $V$ magnitude, with 
$V_{continuum}-V_{total}=0.55$, $0.92$, $1.28$, and $1.26$~mag
on 1998 July 5 and August 30, and 1999 May 5 and July 6, respectively.

\section{Discussion}

\subsection{The two periods in the LC of V4633~Sgr}
The photometric data of the three-year observations of V4633~Sgr
confirms the presence of two independent periodicities in the LC of
V4633~Sgr -- $P_1=3.08$~h $=0.1285$~d and $P_2=3.014$~h
$=0.125576\pm0.000009$~d.

We suggest that $P_2$, is the orbital period of the underlying binary
system as its behaviour
during the three years of photometry is consistent with a stable
period.
In addition, during the photometric monitoring, the waveform of the signal
has maintained its shape. The asymmetric shape of the waveform is
rather unique for orbital modulations, nonetheless, we note its close
similarity to the shape of the orbital modulation of V1974~Cyg in 1996
(Skillman et al. 1997).
The $3.01$~h period is well situated within the range of orbital
periods of cataclysmic variables.
To confirm this suggestion, radial velocity measurements should
be carried out.
In the following we shall consider $P_2$ to be the orbital period,
$P_{\mathrm{orb}}$ of the binary system.

\subsection{The second periodicity}
It is more difficult to interpret the longer period, $P_1$.
This signal is characterized by the following traits:
(1) it is $\sim 2.5$ per cent longer than the binary period.
(2) It is at least an order of magnitude less stable than
$P_{\mathrm{orb}}$, decreasing by $\sim0.3$ per cent during 1999-2000,
with $\dot P\sim -10^{-6}$ (Sec.~\ref{SecPdot}).

Two possible interpretations come to mind. 
One is that the origin of the $P_1$ variation is the rotation of the
white dwarf (WD).
The modulation may arise, for instance, from aspect variation of a hot
spot on or near the surface of the WD.
The small deviation of $P_1$ from the orbital period would then
suggest that V4633~Sgr belongs to the Asynchronous Polars group (BY Cam
stars, hereafter APs).
An alternative interpretation is that the origin of $P_1$ is in an
accretion disc in the system, namely, that it is the period of the
well known phenomenon of superhumps (SH).

In the following two sections we discuss the two interpretations and
some of their implications. The data at our hand seems insufficient
for making a reliable choice between them.

\subsection{Asynchronous Polar interpretation}
APs are a sub-class of magnetic cataclysmic variables,
sharing many of the properties of Polars (AM Her stars), but have a WD
which rotates with a period that differs by $\sim1$ per cent from
the orbital period.
There are four known APs. They are listed in Table~\ref{tableAPs} along
with the major characteristics of their periodicities.
In one AP, V1500~Cyg, the asynchronous rotation is clearly associated
with its nova eruption in 1975.
Two other APs are suggested to have also undergone a recent nova
event (V1432 Aql -- Schmidt \& Stockman 2001;
BY Cam -- Bonnet-Bidaud \& Mouchet 1987).

\begin{table}
  \caption{Periods of Asynchronous Polars. The values for V4633~Sgr are
    given at the bottom for reference.}
  \begin{tabular}{@{}||l|c|c|c|c||@{}}
        & $P_{\mathrm{orb}}$ & $P_{\mathrm{rot}}$ & $\frac{P_{\mathrm{orb}}-P_{\mathrm{rot}}}{P_{\mathrm{orb}}}$ & $\dot P_{\mathrm{rot}}$\\
& h & h &per cent\\ \hline 
{\bf V1500 Cyg}$^{1,2}$  & 3.35 & 3.29 & 1.7 & $4\times10^{-8}$ \\
{\bf BY Cam}$^3$     & 3.35 & 3.32 & 0.9 &  $4\times10^{-9}$\\
{\bf V1432 Aql}$^{4,5}$  & 3.36 & 3.37 & -0.3 &  -$1\times10^{-8}$\\
{\bf CD Ind}$^6$     & 1.85 & 1.83 & 1.1 &  \\\hline 
{\bf V4633 Sgr}$^7$  & 3.01 & 3.08 & -2.3  & $-1.26\times10^{-6}$\\
  \end{tabular}\newline
$^1$Schmidt, Liebert \& Stockman 1995;
$^2$Kaluzny \& Semeniuk 1987;
$^3$Mason~{\it et~al.} 1998;
$^4$Patterson~{\it et~al.} 1995;
$^5$Geckeler \& Staubert 1997;
$^6$Ramsay {\it et~al.} 1999;
$^7$this work.

\label{tableAPs}
\end{table}

The AP interpretation of V4633~Sgr is supported by the monotonous
decrease in $P_1$, the proposed rotation period of the WD
($P_{\mathrm{rot}}$), towards synchronization with
$P_{\mathrm{orb}}$.
A synchronization trend in $P_{\mathrm{rot}}$ is expected in APs due
to the magnetic torque exerted on the WD by the secondary star.
Indeed, such a trend was detected in three of
the four APs (Table~\ref{tableAPs}).
Also, the orbital period of V4633~Sgr, $P_{\mathrm{orb}}=3.01$~h, is
similar to that of three APs (Table~\ref{tableAPs}).
The beat period, $P_3=5.06$~d, detected in 1999 (Sec.~\ref{SecLC99}),
may be naturally explained in the AP framework. If a dipole geometry
is assumed, pole switching is expected to occur at the beat cycle,
modulating the LC at $P_{\mathrm{beat}}$.

However, a simple AP interpretation seems to be inapplicable in
V4633~Sgr due to the following reasons:
$(1)$~The synchronization rate of the proposed $P_{\mathrm{rot}}$ is
$\left | \dot P_1\right |\sim10^{-6}$ -- much larger than in APs:
 $\left |\dot P_{\mathrm{rot}}\right |\sim3\times 10^{-9} - 4\times 10^{-8}$
(Table~\ref{tableAPs}).
$(2)$~In V4633~Sgr, $P_1$ is {\it longer} than $P_{\mathrm{orb}}$, 
while in three of the four APs $P_{\mathrm{rot}}$ is {\it shorter}.
In V1432~Aql, the only AP in which $P_{\mathrm{rot}}>P_{\mathrm{orb}}$, the
difference is marginal.
Even so, the longer $P_{\mathrm{rot}}$ poses some theoretical
difficulties (we note, however, that Schmidt \& Stockman (2001) argue
that $P_{rot}<P_{orb}$ is a possible outcome of a
nova eruption, in slow novae with strong magnetic fields).
Indeed, an alternative model for this object was proposed by Mukai
(1998), in which V1432~Aql is an intermediate polar with a spin period
of $1.12$~h.
$(3)$~The difference between the two periods in V4633~Sgr is $\sim2.3$
per cent --
larger than the value in any of the four APs
(Table~\ref{tableAPs}).
$(4)$~The distinctly asymmetric waveform of $P_{\mathrm{orb}}$ in V4633~Sgr 
is hardly that of an eclipsing system (Sec.~\ref{SecFold}).
If there is no disc in the system, as the AP model suggests, the light modulation
on the orbital period must be ascribed to the 'reflection' effect. Any simple
model of this effect produces symmetric binary LCs. 
$(5)$~If the modulation at $P_{\mathrm{beat}}$ is caused by pole switching, the latter
is expected also to affect $P_1$, invoking a phase shift of
$180^{\circ}$ twice every beat cycle. This effect should reveal
itself both in the PS, reducing the power of the peak associated to
$P_1$, and in the folded LC of $P_1$. However, these effects are not
detected.

Few of the distinctive characteristics of V4633~Sgr may be explained
in the framework of the AP model if they are attributed to short term
changes taking place in the system in the first few years after the
nova outburst. 
Such an irregular behaviour was observed in V1500~Cyg during the first
three years after its outburst. 
These have been described in detail (e.g. Patterson 1979, Lance,
McCall \& Uomoto 1998), and interpreted by Stockman, Schmidt \& Lamb
(1988).

Applying the model of Stockman, Schmidt \& Lamb to V4633~Sgr, we should
assume that the spin of the WD was synchronized with the orbital
revolution prior to the nova event. The rapid expansion of the WD's
envelope during the first stages of the outburst increased the star's
moment of inertia, resulting in a spin-down of the WD by $\gtorder2.5$
per cent.
The decrease in $P_{\mathrm{rot}}$ in 1998 -- 2000 should be attributed to the
contraction of the still expanded envelope of the WD, with the
associated reduction in its moment of inertia.
Thus, $P_{\mathrm{rot}}$ is expected to continue decreasing until the
WD finally regains its original radius.
Following, a slower synchronization trend is expected to occur on the
magnetic synchronization time-scale of the system.
In analogy to V1500~Cyg, if the contraction of the envelope decreases
the moment of inertia of the WD by a magnitude comparable to that
gained during the nova outburst (Patterson 1979, Stockman, Schmidt \&
Lamb 1988), and if the spin acceleration rate maintains its
value of 1999-2000 (Sec.~\ref{SecPdot}), the WD would regain its
pre-nova dimension around the year 2006.

For an order of magnitude calculation, we attribute the change in
$P_1$ in 1998 June -- 2000 entirely to the contraction of the WD's
envelope.
We further assume that the WD is a rigid sphere of mass $M_1$ and
radius $R_1$, rigidly coupled to a thin shell of mass $M_{ph}$ and
radius $R_{ph}$.
Let $\Delta R_{ph}$ and $\Delta \omega$ be the changes in the radius
and the angular velocity of the WD during a time interval $\Delta t$.
Conservation of angular momentum requires that 
$$\frac{2}{3} M_{ph} \left[\left(R_{ph}+\Delta{R_{ph}}\right)^2 -
R_{ph}^2\right] \approx -\frac{2}{5}M_1
R_1^2\frac{\Delta{\omega}}{\omega+\Delta \omega}.$$

Since in 2000 August $R_{ph}\ge R_{1}$, the photosphere radius at
time $\Delta t$ prior to 2000 August is bounded by 
$$R_{ph}\gtorder \sqrt{\frac{3M_1}{5M_{ph}}\frac{\Delta\omega}{\omega+\Delta\omega}}R_1.$$

From the speed class of V4633~Sgr ($t_3\approx 42$~d, Sec.~\ref{SecM0t3})
we infer $M_1\approx1.1$~M$_\odot$ for the mass of the WD (Kato
\& Hachisu 1994). As a rough estimate of the mass of the contracting envelope 
we take $M_{ph}\sim10^{-6}$~M$_\odot$ (Prialnik 1986, Prialnik \& Kovetz
1995).
Inserting these values to the above equation together with the
observed values of $P_1$, we obtain $R_{ph}\gtorder 71R_1$ and $53R_1$
in 1998 June and 1999 May, respectively.

The scenario depicted above is considerably different from the one in
V1500~cyg.
In particular, in the 1975 nova outburst, the WD and its envelope
gained angular momentum through coupling with the orbiting secondary
during the common envelope phase, almost re-synchronizing the WD's spin
with the orbital cycle within a few tens of days after outburst.
This is why in that system, $P_{\mathrm{rot}}$ became shorter than
$P_{\mathrm{orb}}$ as the WD's envelope contracted.
In V4633~Sgr such a coupling either did not take place at all, or was
much less effective in transferring orbital to spin angular momentum.
It is therefore also likely that this system will remain with a spin
period longer than the binary period even after the WD regains its
pre-outburst dimension.

Some other different aspects in the evolution of V4633~Sgr, such as the
apparently larger increase in $P_{\mathrm{rot}}$ during the outburst, and the
longer time-scale of the envelope contraction, may be attributed to a
less massive WD, which is expected to shed more mass
during outburst, and regain its original size on a longer time-scale
(Prialnik \& Kovetz 1995, Kato \& Hachisu 1994).

The AP interpretation should be tested against an observational search for
evidence for the magnetic nature of V4633~Sgr. This should manifest itself, for
example, by strong X-ray radiation and/or circular polarized light, modulated
by the WD rotation period. So far no such observations (or results) have been
reported.\footnote{
Non-detection of linear polarization in 1998 March (Ikeda, Kawabata \& Akitay
2000) is of small relevance, since at that early epoch in the history
of the outburst,  any magnetic polarization would be masked by the
luminous extended photosphere and ejecta. Naturally, as the nova
continues to fade, detection of circular polarization becomes
increasingly feasible.}

\subsection{Permanent superhump interpretation}
Superhumps are periodic brightness variations in the LCs of certain
subgroups of disc-accreting CVs, with a period a few per cent longer than
the orbital period of the binary system (Warner 1995).

Initially, SH were found in the SU~UMa sub-class of
dwarf-novae during superoutburst events.
SH of longer duration, of months and years, are termed 'permanent SH'. 
They appear in LCs of CVs with short orbital periods (typically
$P_{\mathrm{orb}}\ltorder4$~h, Patterson 1999) and high
mass-transfer rates, such as nova remnants, nova-like and AM CVn
systems (for reviews see Patterson 1999, and Retter \& Naylor 2000).
Superhumps also occur in X-ray binaries (e.g. O'Donoghue \& Charles 1996).

The properties of V4633~Sgr make it a good candidate for hosting the SH
phenomenon.
The $3.01$~h orbital period puts V4633~Sgr near the centre of the
the period interval that contains most of the known SH systems
(Patterson, 1998).

The observed stable decline in the brightness of the nova is
consistent with the presence in the system of an accretion disc, that
in the years 1999-2000 is the main source of the optical
luminosity, and which is thermally stable. If mass accretion is indeed
the main luminosity source, we can estimate its rate using
equation~(3) of Retter \& Naylor (2000). In terms of absolute
magnitude it is given by:
$$\dot M_{17} = [10^{(M_V-5.69+\Delta M_i)/-2.5}]\frac{1}{M_1^{4/3}}$$
where $\dot M_{17}$ is the mass-transfer rate in $10^{17}$\,g\,s$^{-1}$,
$M_V$ is the absolute $V$ magnitude of the disc, $M_1$ is the mass of the
WD, and $\Delta M_i=-2.5\log{[(1+1.5\cos{i})\cos{i}]}$ is a correction
to the magnitude due to the inclination angle ($i$) of the disc.
From the $V$-band LC (Fig.~\ref{figLc}), and the estimated
distance and reddening towards V4633~Sgr (Sec.~\ref{SecMMRD}),
we derive $M_{V, 2000}\sim0.5-1.5$.
The non-eclipse shape of the LC (Sec.~\ref{SecFold}) implies that the
inclination angle is $i \le 65^\circ$. For a $M_1\approx 1.1$M$_\odot$
WD, we obtain $\dot M\sim(30 - 300)\times10^{17}$~g/s. 
The critical mass transfer rate, below which the disc is thermally
unstable, is given by Osaki (1996, eq.\,4 therein),
which for $P_{\mathrm{orb}}=3.01$~h takes the value
$\dot M_{crit}\approx1.7\times10^{17}$~g/s. Thus the observed mass
transfer rate in V4633~Sgr is some two orders of magnitude above the
critical value, and the disc is indeed thermally stable.

Superhumps are known to be poor clocks. In permanent
SH systems, the instability in superhump period is:
$\dot P_{SH}=10^{-8} - 5\times10^{-6}$ (Patterson \& Skillman 1994).
The value of $\dot P_1$ that we found in V4633~Sgr in 1999-2000 is
within this range.

The similarity in the shape of the orbital and the superhump
waveforms of V1974~Cyg to those of $P_2$ and $P_1$ (Skillman et
al. 1997) serves as a further support to the SH interpretation.

\begin{figure}
 \centerline{\epsfxsize=3.5in\epsfbox{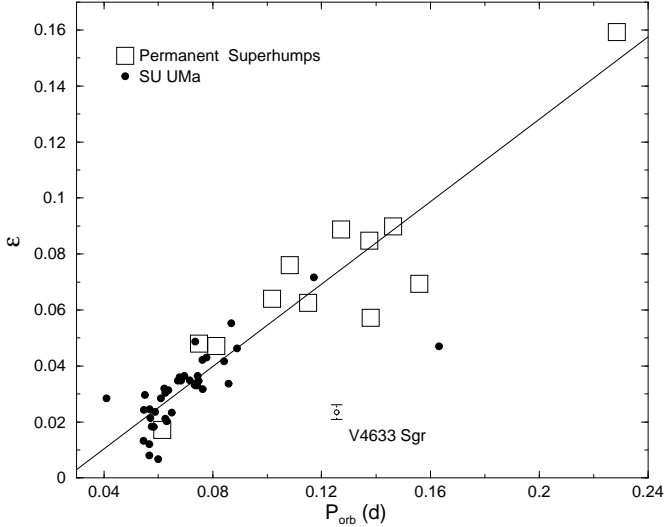}}
  \caption{The Period excess -- $P_{\mathrm{orb}}$ relation of superhump
    systems. Data were taken from Patterson (1998; 1999),
    and Retter et al. (2001)
    The solid line is a linear fit to the data.}
 \label{figSSrel}
 \end{figure}

On the weak side of the SH interpretation stands the value of the
period excess $\epsilon\equiv (P_{SH}-P_{\mathrm{orb}})/P_{\mathrm{orb}}$.
Superhump systems are known to follow a nearly linear relation between
$\epsilon$ and
$P_{\mathrm{orb}}$ (Stolz \& Schoembs 1984).
In V4633~Sgr the measured value, $\epsilon=0.024\pm0.003$, is about
a third of the value expected for $P=3.01$~h (Fig.~\ref{figSSrel}).
Inspection of Fig.~\ref{figSSrel} reveals, however, that while the
point of V4633~Sgr deviates the most from the empirical linear
relation, it is not qualitatively exceptional. All other
points representing permanent SH systems (empty squares symbols) are
distributed with a rather large scatter around the regression line. Two of
them, V603~Aql and BH~Lyn, show an exceptionally large deviation
relative to the other systems.
We should also note that the apparent large deviation of the black dot
representing the SU~UMa system CN~Ori should be treated with caution,
until its period excess is confirmed by further observations
(Patterson, private communication).

Since the disc precession is caused by the perturbation of the
secondary star, the precession rate should be proportional to the
secondary's mass, $M_2$. 
Such a relation was found by Osaki (1985), who examined the motion of
a free particle in a binary potential.
In particular, for a disc with radius $\approx 0.46$ the binary
separation (this is approximately the disc radius at the 3:1
resonance where SH are most likely to occur), Osaki derived the
relation:
$$P_{\mathrm{orb}}/P_{\mathrm{beat}}~\approx~0.233~\frac{q}{\sqrt{1+q}},$$
where $q~\equiv~M_2/M_1$.
For V4633~Sgr, this relation yields $q~\approx~0.10-0.11$.
Since the mass of the WD should be smaller than the Chandrasekhar
mass ($1.44$~M$_\odot$), the mass of the secondary is bounded by
$M_2~\ltorder~0.16$~M$_\odot$.

On the other hand, if the secondary is a Roche-lobe filling, main-sequence
star, its mass can be derived analytically (e.g. Warner 1995), if
$P_{\mathrm{orb}}$ is known. 
An Empirical $P_{\mathrm{orb}}-M_2$ relation yields a result similar to the
analytical ones (Smith \& Dhillon 1998).
For $P_{\mathrm{orb}}=3.01$ hr, the mass of a main-sequence secondary is
$M_2\approx0.27$~M$_\odot$, much larger than the limit obtained above.

This inconsistency may infer that the cause for the exceptionally
small $\epsilon$ may be an under-massed secondary star, which is off
the main sequence.
In this case, V4633~Sgr may be an extremely evolved CV system
(e.g. Howell, Rappaport \& Politano 1997; Patterson 1998).

The $5.06$~d signal observed in 1999 ($P_3$, Sec.~\ref{SecLC99}),
presents another difficulty for the permanent SH scenario.
This period corresponds to the beat
period between $P_{\mathrm{orb}}$ and $P_{\mathrm{SH}}$.
It is therefore natural to interpret
this signal as arising from the precession of the accretion disc.
However, theoretically, apsidal precession of an
eccentric disc is not expected to modulate the light of the nova
(Skillman \& Patterson 1993; Patterson 1998). 
We note however that such modulations were actually observed in the
permanent SH system AH~Men (H 0551-819) in 1993-4, when the
object showed positive SH (Patterson 1995).

The permanent SH interpretation may be tested photometrically during
the next few years. Superhump periods are found to wander
about a mean value, and therefore $\dot P_{\mathrm{SH}}$ is expected
to occasionally change its sign.

\subsection{Further parameters of V4633~Sgr}
\subsubsection{The visual light curve}
\label{SecM0t3}
We derive some of the properties of the visual LC of V4633~Sgr
using magnitudes of the nova published in the IAU Circulars and in the
VSNET web site, and the LC presented by Liller \& Jones (1999).

The data suggest that the nova was discovered before reaching
maximum brightness, as was already pointed out by Liller \& Jones
(1999).
However, the scatter in magnitude estimates during the
first few days after discovery, does not allow us to determine the
exact timing and magnitude of maximum brightness.
We can only conclude that the nova reached maximum light sometime
between JD~2450895.5 and JD~2450898.5.
We adopt the value of $m_{v,0}=7.7\pm0.1$ for its visual magnitude at
maximum. 

From the VSNET data we estimate decline rates of
$t_{2,v}=19\pm3$~d, and $t_{3,v}=42\pm5$~d,
somewhat longer than the estimation of Liller \& Jones (1999) --
$t_{3,v}\approx35$~d.

By the classification scheme of Duerbeck's (1981), V4633~Sgr should be
classified as $Ba$-type nova -- moderately fast with minor irregular
fluctuations during decline.

\subsubsection{Photometric changes in 1998 June -- August}
Around 1998 June, there was an apparent bend in the visual
and $V$ LCs (Sec.~\ref{SecLC}, Fig.~\ref{figLc}).
Leibowitz (1993) noted that such a feature is found in the visual LCs of
many classical novae, and suggested to attribute it to the decay of
the WD's light level below the brightness emitted by the
accreted material.
This interpretation was cast into quantitative form in models
suggested recently by Hachisu et al. (2000) and by Hachisu \&
Kato (2000) for the LCs of the two recurrent novae V394~CrA and U~Sco.

Shortly after the change in the slope of the LC, in 1998 July -- August,
the $I$ LC deviated from its smooth decline, forming an apparent bump.
A similar bump was seen in the $B$ and $V$ LCs
(Sec.~\ref{SecLC}, fig.~\ref{figLc}, inset frame).
Two of our spectra, taken at the same time, on 1998 July 5 and August
30, show the emergence of strong {[O\,{\sc iii}]}~$\lambda\lambda 4959,5007$
emission lines. (Sec.~\ref{SecSpec}).
The simultaneous occurrence of the two effects was observed in
a few other novae, and was connected with the beginning of the
nebular stage (Chochol et al. 1993).

During July -- August another photometric peculiarity occurred -- the
LC was modulated in a different form than previously.
In particular, the periodicities of 1998 June were not detected during
these months (Sec. \ref{SecAug98}).
We offer no explanation for this phenomenon, or to its possible
connection to the aforementioned phenomena.

\subsubsection{Interstellar reddening}
We can estimate the interstellar reddening towards V4633~Sgr in three
ways.
First, we consider the observed Balmer Decrement in the spectra of
the nova
(in the following we neglect the  contribution of the
{[N\,{\sc ii}]} $\lambda\lambda 6548,6584$ lines to the measured
H$\alpha$ line-intensity. From the {[O\,{\sc iii}]}~$(5007+4959)/4363$
line-ratio and the {[N\,{\sc ii}]}~$5755$ line-intensity
(Osterbrock 1989), we estimate it to be less than 5 per cent of the
measured flux).
Slightly more than three months after maximum, the line intensity ratio
H$\alpha$/H$\beta$ was as high as $6.8$, probably due to self
absorption (Williams 1994).
Our spectra show the progressive decrease of this line
ratio during the following year. Between our last two spectroscopic
observations the trend of decrease has flattened considerably. About 15 months
after outburst, in our last spectrum measurement, this line ratio reached the
value $3.52\pm0.10$  (Table~\ref{tableSpec}).
.Attributing the difference between this value and the theoretical case B
value of $2.8$ (Osterbrock 1989) entirely to dust extinction, and using the
numerical form of the Whitford (1958) reddening curve
given by Miller \& Mathews (1972), we obtain a reddening of
$E(B-V)=0.21\pm0.03$. 

A second way to estimate the reddening is from the He triplet ratio,
$5876/4471=2.9$, which seems to be insensitive to radiation transfer
effects (Ferland 1977). 
The observed ratio on August 30, 1998 was $3.7\pm0.4$, leading to
$E(B-V)\sim0.23$, in agreement with the value derived from the H
lines.
We did not measure this line ratio in the spectrum of 1998 July 5,
because the uncertainty in the measurement of the He~$\lambda 4471$
line was much larger at that epoch.
We note that, as pointed out by Ferland (1977), this method is
inaccurate due to the small baseline and also to the weakness of the
He~$\lambda 4471$ line.

We also estimated the reddening using colour photometry of the nova
shortly after outburst.
Novae have intrinsic colours $(B-V)_0=+0.25\pm0.05$ at
maximum (Downes \& Duerbeck 2000), and $(B-V)_0=-0.02\pm0.04$ at
$2$~mag below maximum light (van den Bergh \& Younger 1987).
Observations in the $B$ and $V$ bands by S. Kiyota, reported in VSNET,
yield a colour index $(B-V)=+0.50$ on 1998 March 25, slightly below
maximum light, and $(B-V)=+0.24$ on 1998 April 19, slightly below
maximum plus $2$~mag. At these two dates, the intrinsic colour of the
nova was somewhat redder than the corresponding two 'standard' values
quoted above. 
The difference between the two pairs of values constrains the
interstellar reddening towards the nova to $E(B-V)\ltorder 0.25$, in
agreement with the value derived from spectroscopy.

The small degree of reddening towards V4633~Sgr was noticed also by
Rudy et al. (2000).
The survey of Neckel and Klare (1980) confirms the relatively low
extinction towards V4633~Sgr ($A_V\approx0.65$, if we use
$R=A_V/E(B-V)=3.1$).
The galactic coordinates of V4633~Sgr are $(l,b)=(5.128,\ -6.231)$, and
its estimated distance is $d\approx 9$~kpc (Sec.~\ref{SecMMRD}).
The corresponding field in Neckel and Klare (1980) is $237$, for which
they found $A_v\approx 0.6-1.1$ at distances of $5-8$~kpc.

\subsubsection{Maximum magnitude and distance}
\label{SecMMRD}
We estimate the absolute magnitude of V4633~Sgr at maximum
brightness, $M_{V,0}$, by two methods. First, we use
the empirical maximum magnitude -- rate of decline (MMRD) relation
obeyed by novae.
We use the linear MMRD relations for $t_2$ and $t_3$ derived by Downes \&
Duerbeck (2000) from an ensemble of 28 measured novae. Their relations
yield for V4633~Sgr values of $-8.1\pm0.6$ and
$M_{V,0}=-7.9\pm0.8$~mag, respectively.
Downes \& Duerbeck (2000) also derived MMRD relations for $t_2$ and
$t_3$ from 17 nova classified as $B$, $C$ and $D$ in the LC
classification scheme of Duerbeck (2000).
These relations yield for V4633~Sgr values of $-7.4\pm1.1$ and
$M_{V,0}=-7.2\pm1.7$~mag, respectively. 

We can also estimate $M_{V,0}$ using 
the absolute magnitude 15 days after maximum, which appears to be
independent of speed class (Warner 1995).
Downes \& Duerbeck (2000) derived from 28
objects a value of $M_{V,15}=-6.05\pm0.44$~mag.
This value, together
with our estimate of the visual magnitude of V4633~Sgr at maximum,
$m_{v,0}=7.7\pm0.1$~mag, (Sec.~\ref{SecM0t3}), and with the value of
$m_{V,15}=9.45\pm0.06$~mag which was measured for  V4633~Sgr at WO on
JD~2450912.51, yields $M_{V,0}=-7.8\pm0.5$~mag for the absolute
magnitude of V4633~Sgr at maximum brightness.

We adopt the average of the above results, $M_{V,0}\approx-7.7$~mag
for the intrinsic magnitude at maximum.

Incorporating our estimations of the reddening, and the intrinsic and
apparent maximum brightness of the nova into the distance modulus equation
(Allen 1976),
we derive a distance of $8.9\pm2.5$~kpc to V4633~Sgr, compatible 
with the estimation of Ikeda, Kawabata \& Akitay (2000).
We note, that the derived distance to V4633~Sgr implies that it
probably belongs to the population of ``bulge'' novae.
Indeed, the spectroscopic classification of V4633~Sgr as a 
Fe\,{\sc ii} nova, as  well as its rate of decline, are
characteristic of ``bulge'' novae (Della~Valle \& Livio 1998).

\section{Summary}
Three-year observations of V4633~Sgr revealed two photometric
periodicities in the light curve of the nova.
We interpret the shorter one, $P_2=3.014$~h as the orbital period of
the underlying binary system.
The longer period, $P_1=3.08$~h, varied during 1998-2000 with 
$\dot P_1=(-1.26\pm0.05)\times 10^{-6}$. 
The beat of the two periods, $P_3=5.06$~d, was probably present in the
LC in 1999.

The period $P_1$ may be interpreted as a permanent superhump, or alternatively,
as the spin period of the white dwarf in a nearly synchronous magnetic system.
V4633~Sgr would be a unique SH system, since its relative period excess
is exceptionally small -- $\sim2.5$ per cent.
This may imply an extremely low mass ratio.
The characteristics of V4633~Sgr are also unique for the
near-synchronous polar model.

Further photometric monitoring of the V4633~Sgr in the next few years 
will probably allow us to determine the classification of the system,
since the non-orbital period is expected to evolve differently in the
two models.
Radial velocity measurements should be done to confirm the orbital period.
Time-resolved polarimetry and X-ray observations should be conducted
to test the near synchronous polar interpretation.

\section*{Acknowledgments}
We are grateful to Dina Prialnik for some very useful discussions.
This research has made use of the VSNET database.
We thank Albert Jones for sending his visual estimates to us.
Astronomy at the WO is supported by grants from the Israel Science
Foundation. AR was supported by PPARC during most of the period this
work was carried out, and is currently supported by the Australian
Research Council.

\label{lastpage}

\cleardoublepage

\section*{Appendix A. Log of observations}

\begin{tabular}{ll}
\\
  \begin{tabular}{@{}ccccccc@{}}
UT&Time of Start&Run Time&\multicolumn{3}{c}{Points per Filter}\\
   Date  &(HJD-2450900) & (hours)& I & V & B &\\
\\ 

980408 &  12.51 & 0.04  & 1     & 1     & 1\\
980409 &  13.51 & 2.3   & 569   &       & \\
980412 &  16.51 & 2.1   & 272   & 2     & \\
980413 &  17.50 & 2.4   & 163   & 164   & \\
980420 &  24.50 & 2.1   & 104   & 99    & \\
980509 &  43.47 & 0.7   & 43    &       & \\
980511 &  45.44 & 3.6   & 214   & 168   & \\
980512 &  46.41 & 1.3   & 89    & 88    & \\
980607 &  72.35 & 5.5   & 493   &       & \\
980608 &  73.35 & 5.5   & 284   & 280   & 1\\
980609 &  74.34 & 5.3   & 211   & 210   & \\
980611 &  76.34 & 5.5   & 243   & 242   & \\
980612 &  77.35 & 5.4   & 452   &       & \\
980613 &  78.38 & 4.7   & 173   &       & \\
980703 &  98.27 & 4.2   & 57    & 57    & 54\\
980704 &  99.28 & 4.2   & 45    & 45    & 45\\
980705 & 100.38 & fosc  &       &       & \\
980709 & 104.26 & 5.7   & 325   &       & \\
980710 & 105.26 & 5.8   & 38    & 38    & 37\\
980711 & 106.24 & 4.2   & 38    & 38    & 37\\
980724 & 119.38 & 2.2   & 15    & 15    & 12\\
980729 & 124.34 & 2.2   & 16    & 15    & 15\\
980730 & 125.31 & 2.7   & 20    & 20    & 19\\
980803 & 129.36 & 1.9   & 15    & 15    & 15\\
980806 & 132.23 & 4.5   & 36    & 36    & 34\\
980807 & 133.23 & 4.4   & 30    & 30    & 30\\
980808 & 134.23 & 4.5   & 33    & 32    & 32\\
980809 & 135.25 & 4.0   & 30    & 29    & 29\\
980810 & 136.24 & 4.3   & 38    & 38    & 36\\
980811 & 137.28 & 2.5   & 17    & 17    & 17\\
980830 & 156.25 & fosc  &       &       & \\
980911 & 168.21 & 0.6   & 2     & 2     & 2\\
980913 & 170.21 & 0.7   & 3     & 2     & 2\\
980918 & 175.20 & 0.3   & 1     & 1     & 1\\
980919 & 176.20 & 0.3   & 2     & 1     & 1\\
980920 & 177.21 & 0.3   & 1     & 1     & 1\\
990418 & 387.57 & 0.1   &       & 1     & \\
990425 & 394.46 & 3.3   & 24    & 22    & 21\\
990426 & 395.45 & 3.7   & 28    & 24    & 23\\
990427 & 396.45 & 3.6   & 28    & 23    & 23\\
990428 & 397.45 & 3.5   & 24    & 24    & 24\\
990429 & 398.44 & 2.9   & 19    & 19    & 18\\
990502 & 401.52 & fosc  &       &       & \\
990503 & 402.44 & 3.7   & 34    & 34    & 1\\
990518 & 417.39 & 4.8   & 35    & 32    & 31\\
990519 & 418.53 & 1.0   & 7     & 7     & 7\\
990520 & 419.53 & 1.0   & 7     & 7     & 7\\
990523 & 422.37 & 5.1   & 35    & 33    & 32\\
990524 & 423.37 & 5.2   & 39    & 33    & 32\\
990525 & 424.36 & 5.3   & 39    & 34    & 31\\
990526 & 425.41 & 4.2   & 136   & 1     & 1\\
  \end{tabular}

&
\begin{tabular}{@{}ccccccc@{}}
UT&Time of Start&Run Time&\multicolumn{3}{c}{Points per Filter}\\
   Date  &(HJD-2450900) & (hours)& I & V & B\\
\\ 
990528 & 427.35 & 4.2   & 40    & 4     & 2\\
990529 & 428.35 & 5.6   & 50    & 42    & 3\\
990615 & 445.44 & 0.02  & 1     &       & \\
990619 & 449.40 & 0.7   & 5     & 5     & 5\\
990621 & 451.41 & 0.1   & 1     & 1     & 1\\
990622 & 452.43 & 0.1   & 1     & 1     & 1\\
990624 & 454.28 & 6.7   & 44    & 27    & 18\\
990625 & 455.28 & 6.6   & 52    & 43    & \\
990626 & 456.27 & 6.9   & 74    & 43    & \\
990630 & 460.27 & 6.7   & 71    & 56    & \\
990701 & 461.27 & 6.6   & 73    & 63    & \\
990702 & 462.28 & 6.4   & 51    & 44    & 37\\
990703 & 463.26 & 6.8   & 109   & 97    & 0\\
990704 & 464.26 & 6.8   & 108   & 101   & 0\\
990705 & 465.25 & 6.6   & 56    & 44    & 32\\
990706 & 466.29 & fosc  &       &       & \\
990707 & 467.24 & 7.1   & 117   & 101   & \\
990729 & 489.26 & 5.0   & 241   &       & \\
990730 & 490.22 & 5.8   & 156   &       & \\
990731 & 491.22 & 56.0  & 154   &       & \\
990801 & 492.22 & 6.1   & 155   &       & \\
990802 & 493.22 & 0.3   & 6     &       & \\
990803 & 494.23 & 4.0   & 101   &       & \\
000327 & 731.60 & 0.2   & 1   & 1  & 1  \\
000416 & 751.47 & 2.7   & 26  &    &    \\
000417 & 752.46 & 3.4   & 44  & 1  & 2  \\
000421 & 756.45 & 1.8   & 24  &    &    \\
000422 & 757.46 & 3.1   & 2   & 29 &    \\
000511 & 776.41 & 4.3   & 28  & 28 & 1  \\
000513 & 778.41 & 3.8   & 29  & 28 & 1  \\
000514 & 779.44 & 3.1   & 27  & 24 & 1  \\
000515 & 780.43 & 2.6   & 22  & 22 & 2  \\
000720 & 846.35 & 2.6   & 43  &    &    \\
000721 & 847.23 & 6.2   & 111 &    &    \\
000722 & 848.23 & 6.2   & 111 &    &    \\
000723 & 849.23 & 6.1   & 109 &    &    \\
000727 & 853.37 & 0.9   & 16  &    &    \\
000728 & 854.31 & 2.6   & 48  &    &    \\
000729 & 855.30 & 2.6   & 42  &    &    \\
000803 & 860.31 & 2.1   & 33  &    &    \\
000804 & 861.34 & 2.4   & 42  &    &    \\
000804 & 861.24 & 2.5   & \multicolumn{3}{l}{\ (529 'clear')}\\
000805 & 862.24 & 5.2   & 94  &    &    \\
000806 & 863.22 & 5.5   & 93  &    &    \\
000817 & 874.27 & 2.9   & 40  &    &    \\
000818 & 875.22 & 4.2   & 72  &    &    \\
000819 & 876.21 & 4.6   & 84  &    &    \\
000820 & 877.21 & 4.9   & 629 &    &    \\ 
000821 & 878.21 & 4.8   & 686 &    &    \\  
000822 & 879.21 & 2.1   & 29  &    &    \\
\\
  \end{tabular}
\end{tabular}


\begin{thebibliography}{99}

\bibitem[]{} Allen C. W., 1976, Astrophysical Quantities, The Athlone
  Press, University of London

\bibitem[]{} Bonnet-Bidaud J. M., Mouchet M., 1987, A\&A, 188, 89

\bibitem[]{} Brosch N., Goldberg Y., 1994, MNRAS, 268, L27

\bibitem[]{} Chochol D., Hric L., Urban Z., Kom\v{z}\'{\i}k R., Grygar J.,
  Papou\v{s}ek J., 1993. A\&A, 277, 103

\bibitem[]{} Della Valle M., Livio M., 1998, ApJ, 506, 818


\bibitem[]{} Della Valle M., Pizzella A., Bernardi M., 1998, IAU Circ., 6848

\bibitem[]{} Downes R. A., Duerbeck H. W., 2000, AJ, 120, 2007

\bibitem[]{} Duerbeck H. W., 1981, PASP, 93, 165

\bibitem[]{} Efron B., Tibshirani R. J., 1993, An Introduction to the
  Bootstrap. Chapman \& Hall

\bibitem[]{} Ferland G.J.,1977, ApJ, 215, 873

\bibitem[]{} Geckeler R. D., Staubert R., 1997, A\&A 325 1070

\bibitem[]{} Hachisu I., Kato M.,2000,  ApJ 540, 447

\bibitem[]{} Hachisu I., Kato M., Kato T.,Matsumoto K., Ken'ichi N.,
  2000, ApJ, 534, L189
\bibitem[]{} Howell S. B., Rappaport S., Politano M., 1997, MNRAS, 287,
  929

\bibitem[]{} Ikeda Y., Kawabata K. S., Akitaya H., 2000, A\&A, 355, 256

\bibitem[]{} Jones, A. F., 1998, IAU Circ., 6848

\bibitem[]{}Kaluzny J., Semeniuk I., 1987, Acta Astron., 37, 349

\bibitem[]{} Kaspi S., Ibbetson P. A., Mashal E., Brosch N., 1995,
  Wise Obs. Tech. Rep., No 6

\bibitem[]{} Kaspi S., Smith. P.S., Netzer H., Maoz D., Jannuzi B.T.,
  Giveon U., 2000, ApJ, 533, 631

\bibitem[]{} Kato M., Hachisu I., 1994 ApJ, 437, 802

\bibitem[]{} Lance C. M., McCall M. L., Uomoto A. K., 1988, ApJS, 66, 151

\bibitem[]{} Leibowitz E.M., 1993, ApJ, 411, L29

\bibitem[]{} Leibowitz, E. M., Ibbetson, P., Ofek, E. O., 1999, BaltA, 9, 403

\bibitem[]{} Liller W., 1998, IAU Circ., 6846

\bibitem[]{} Liller W., Jones A. F., 1999, IBVS, 4664

\bibitem[]{} Lipkin Y., Leibowitz E. M., 2000, IAU Circ., 7372

\bibitem[]{} Lipkin Y., Retter A., Leibowitz E. M., 1998, IAU Circ., 6963

\bibitem[]{} Mason P. A., Ramsay G.; Andronov I., Kolesnikov S.,
  Shakhovskoy N., Pavlenko E. 1998, MNRAS, 295, 51

\bibitem[]{} Miller J. S., Mathews W. G., 1972, ApJ, 172, 593

\bibitem[]{} Mukai K., 1998 ApJ, 498, 394

\bibitem[]{} Netzer H., Heller A., Loinger F., Alexander T., Baldwin
  J. A., Wills B.J., Han M., Frueh M., Higdon L., 1996, MNRAS, 279, 429

\bibitem[]{} Neckel Th., Klare G., 1980, A\&AS, 42, 251

\bibitem[]{} O'Donoghue D, Charles P. A. 1996, MNRAS, 282, 191

\bibitem[]{} Osaki Y., 1985,A\&A, 144, 369

\bibitem[]{} Osaki Y., 1996 PASP, 108, 39

\bibitem[]{} Osterbrock D.E. 1989, Astrophysics of Gaseous Nebulae and
  Active Galactic Nuclei, University Science Books,Mill Valley, California

\bibitem[]{} Patterson J., 1979, ApJ, 231, 789

\bibitem[]{} Patterson J., 1995, PASP, 107, 657

\bibitem[]{} Patterson J., 1998, PASP, 110, 1132

\bibitem[]{} Patterson J., 1999, in ``Disk Instabilities in Close Binary
  Systems. 25 Years of the Disk-Instability Model'', Eds. S. Mineshige,
  J. C. Wheeler. Universal Academy Press, Inc., p. 61

\bibitem[]{} Patterson J., Skillman D. R., 1994, PASP, 106, 1141

\bibitem[]{} Patterson J., Thomas G., Skillman D. R., Diaz M., 1993,
  ApJS, 86, 235 

\bibitem[]{} Patterson J., Skillman D. R., Thorstensen J., Hellier C.,
  1995, PASP, 107, 307

\bibitem[]{} Prialnik D., 1986, ApJ, 310, 222

\bibitem[]{} Prialnik D., Kovetz A., 1995, ApJ, 445, 789

\bibitem[]{} Ramsay G., Buckley D. A. H., Cropper M., Harrop-Allin,
  M. K., 1999, MNRAS, 303, 96

\bibitem[]{} Retter A., Naylor T., 2000, MNRAS, 319, 510

\bibitem[]{} Retter A., Leibowitz E. M., Kovo-Kariti O., 1998, MNRAS,
  293, 145

\bibitem[]{} Retter A., Hellier C., Augusteijn T., Naylor T., Bembrick
  C., McCormick J., Velthuis F., 2001, MNRAS, submitted

\bibitem[]{} Rudy R. J., Lynch D. K., Mazuk S., Puetter R. C.,
  Woodward C. E., 1999, IAU Circ., 7259

\bibitem[]{} Rudy R. J., Lynch D. K., Mazuk S., Venturini C., Puetter
  R. C., Armstrong T., 2000, IAU Circ., 7491

\bibitem[]{} Scargle J.D., 1982, ApJ, 263, 835

\bibitem[]{} Schmidt G. D., Stockman H. S., 2001, ApJ, 548, 410

\bibitem[]{} Schmidt G. D., Liebert J., Stockman H. S., 1995, ApJ, 441, 414

\bibitem[]{} Skiff, B. A., 1998, IAU Circ., 6851

\bibitem[]{} Skillman D. R., Patterson J., 1993, ApJ, 417, 298

\bibitem[]{} Skillman D. R., Harvey D.,Patterson J., Vanmunster T.,
  1997, PASP, 109, 114

\bibitem[]{} Smith D. A., Dhillon V. S., 1998, MNRAS, 301, 767

\bibitem[]{} Stetson P. B., 1987, PASP, 99, 191

\bibitem[]{} Stockman H. S., Schmidt G. D., Lamb D. Q., 1988, ApJ, 332, 282

\bibitem[]{} Stolz B., Schoembs R., 1984, A\&A, 132, 187
             
\bibitem[]{} van den Bergh S., Younger P. F., 1987, A\&AS, 70, 125

\bibitem[]{} Warner B., 1995, Cataclysmic Variable Stars. Cambridge
  University Press

\bibitem[]{} Whitford A. E., 1958, AJ, 63, 201

\bibitem[]{} Williams R. E., 1994, ApJ, 426, 279

\bibitem[]{} Williams R. E., Hamuy M., Phillips M. M., Heathcote
 S. R., Wells L., Navarrete M., 1991, ApJ, 376, 721

\bibitem[]{} Williams R. E., Phillips M. M., Hamuy M., 1994, ApJS, 90, 297


\end{thebibliography}
\end{document}